\documentclass[preprint]{revtex4}%
\usepackage{enumerate}
\usepackage{graphicx}
\usepackage{dcolumn}
\usepackage{bm}
\usepackage{amsmath}
\usepackage{amsxtra}
\usepackage{amstext}
\usepackage{amssymb}
\usepackage{latexsym}
\usepackage{amscd,verbatim}
\usepackage[all]{xy}%
\usepackage{amsfonts}
\providecommand{\U}[1]{\protect\rule{.1in}{.1in}}

\begin{document}
\title{ {Determination of the Dynamic ITER Energy Confinement Time Scalings} }
\author{Giorgio SONNINO$^{1,}{}^{2\star}$, Alberto SONNINO$^{3}$, Jarah EVSLIN$^{4}$,
Gy$\ddot{\mathrm{o}}$rgy STEINBRECHER$^{5}$}
\affiliation{${}^{1}$Universit{\'e} Libre de Bruxelles (U.L.B.), Department of Theoretical
Physics and Mathematics Campus Plaine C.P. 231 Brussels - Belgium. }
\affiliation{${}^{2}$Royal Military School (RMS), Av. de la Renaissance 30 1000 Brussels -
Belgium. }
\affiliation{${}^{3}$Ecole Polytechnique de Louvain (EPL), Universit{\'e} Catholique de
Louvain (UCL) Rue Archim$\grave{\mathrm{e}}$de 1 bte L6.11.01, 1348
Louvain-la-Neuve - Belgium. }
\affiliation{${}^{4}$ High Energy Nuclear Physics Group, Institute of Modern Physics,
Chinese Academy of Sciences, Lanzhou - China}
\affiliation{${}^{5}$Physics Department - University of Craiova Str. A. I. Cuza 13 200585
Craiova - Romania. }

\begin{abstract}
We derive the differential equation, which is satisfied by the ITER scalings
for the dynamic energy confinement time. We show that this differential equation can also
be obtained from the differential equation for the energy confinement time,
derived from the energy balance equation, when the plasma is near the steady
state. We find that the values of the scaling parameters are linked to the
second derivative of the power loss, estimated at the steady state. As an example of an application, the
solution of the differential equation for the energy confinement time is
compared with the profile obtained by solving numerically the balance
equations (closed by a transport model) for a concrete Tokamak-plasma.
\vskip 0.5truecm \noindent PACS numbers: 28.52.-s, 28.52.Av

\noindent${}^{\star}$ Email: gsonnino@ulb.ac.be

\end{abstract}
\maketitle








\section{Introduction}

\label{I}

Global scaling expressions for the energy confinement time, $\tau_{E}$, or the
stored energy, $W$, are powerful tools for predicting the confinement
performance of burning plasmas \cite{kardaun1}, \cite{tsunematsu},
\cite{doyle}. The fusion performance of ITER is predicted using three
different techniques: statistical analysis of the global energy confinement
data in the parameters (simple (multivariate) linear regression tools can be
used to determine the parameters from a set of data) \cite{wagner},
\cite{ITER}, a dimensionless scaling analysis, based on dimensionless physics
parameters \cite{luce}, \cite{ITER}, \cite{Kadomtsev}, and theory -based on
transport models and modelling the plasma profiles \cite{kritz},
\cite{weiland} and \cite{bateman}. Although the three methods give overlapping
predictions for the performance of ITER, the confidence interval of all of the
techniques is still quite wide \cite{cordey}. The Confinement Database and
Modelling Expert Group recommended for ITER design the so-called
$ITERH-98P(y,2)$ confinement scaling \cite{ITER}, \cite{fusionwiki}:
\begin{equation}
\label{I1}\tau_{E}^{H98(y,2)}=0.0562\ I_{p}^{ 0.93}R^{1.97}\epsilon
^{0.58}\kappa^{0.78}B_{0\phi}^{0.15} {\bar n}_{e}^{0.41} P^{-0.69}M^{0.19}%
\end{equation}
\noindent Here, the parameters are the plasma current $I_{p}$, the major
radius $R$, the inverse aspect ratio $\epsilon= a/R$ (with $a$ denoting the
minor radius of the Tokamak), the elongation $\kappa$, the toroidal magnetic
field (at the major radius $R$) $B_{0\phi}$, the central line averaged
electron density ${\bar n}_{e}$, the loss power $P$, and the ion mass number
$M$, respectively. The expression (\ref{I1}) is valid for the ELMy H-mode
thermal energy confinement time. The $2\log$-linear interval was determined to
be $20\%$. By recent analyzing the enlarged $ITERH.DB3$ dataset, the practical
reliability of the $ITERH-98(y,2)$ scaling was confirmed and $2\log$-linear
interval was reduced to $14\%$ \cite{kardaun}. Tables showing some of the most
generally used sets of scaling parameters for the ELMy H-mode and L-mode can
be found in Refs \cite{ITER}, \cite{cordey1}, \cite{yushmanov} and \cite{kaye}.

\noindent For stellarators, a similar scaling has been obtained
\cite{dinklage}, \cite{yamada}
\begin{equation}
\label{I2}\tau_{E}=0.148\ R^{0.64}a^{2.33}{\bar n}_{e}^{0.55} B_{0\phi}
^{0.85}\iota^{0.41}P^{-0.61}
\end{equation}
\noindent where $\iota/2\pi$ is the rotational transform (or the field line pitch).

\noindent The confinement time is defined as
\begin{equation}\label{I2a}
\tau_{E}=\frac{W_{e}}{P_{tot}-{\dot{W}}_{e}}=\frac{W_{e}}{P_{Q}}
\end{equation}
\noindent where $W_{e}$, $P_{Q}$ and $P_{tot}$ are the internal energy, the
power loss and the power source, respectively. From Eq.(\ref{I2a}) results
that when the tokamak is not in the steady state the quantity $\tau_{E}$ is a
time dependent quantity. Hence, $\tau_{E}$, given by Eqs~(\ref{I1}) and
(\ref{I2}), is viewed as a time-dependent variable, which depends on a
collection of variables \textit{dependent} on time (\textit{e.g.}, ${\bar n}$,
$P$, etc). The value of $\tau_{E}$ at the \textit{steady state condition}
${\dot\tau}_{E}=0$, \textit{attained at some time moment} $t_{0}$, corresponds
to the numerical value provided by the database. For example, the point
prediction for the thermal energy confinement time in ITER is ($\tau_{E}$,
${\dot\tau}_{E}$) = ($3.6\ sec, 0$).

\noindent The main objective of this work is to estimate the energy
confinement time, close to the steady state. $\tau_{E}$ at the steady state
condition is calculated by using the expression
\begin{equation}\label{I2a1}
\tau_{E}^{0}= \frac{W_{estat.}}{P_{Qstat.}}
\end{equation}
\noindent where $W_{estat.}$ and $P_{Qstat.}$ are obtained by solving the
\textit{stationary} balance equations. An example of calculation can be found
in Ref.~\cite{cardinali}. To estimate the dynamic confinement time we should
solve the evolutive balance equations. However, this is a very complex task.
An alternative strategy (which is the one that we shall adopt here) consists
in deriving the time differential equation for the energy confinement time,
with $\tau_{E}^{0}$, estimated by using Eq.~(\ref{I2a1}), playing the role of
the initial condition. We show that $\tau_{E}$ is the solution of a nonlinear
differential equation of second order in time, obtained by combining
Eq.~(\ref{I2a}) with the (dynamic) balance equations. The critical fact which makes our approach useful is that in the vicinity of the
stationary state, this differential equation depends only on one coefficient
which varies \textit{very slowing in time}
\begin{equation}\label{I2a2}
\left\{
\begin{array}
[c]{ll}
& \!\!\!\!\tau_{E}{\ddot\tau}_{E}-{\dot\tau}_{E}^{2}=\chi(t)\tau_{E}^{2}\\
& \\
& \!\!\!\!\tau_{E}^{0}=\frac{W_{{estat.}_{{}_{}}}}{P_{Qstat.}}\quad;\quad{\dot\tau}
_{E}=0
\end{array}
\right.
\end{equation}
\noindent where $\chi(t)\simeq \chi_0(t-t_0)$, and $\chi_0$ is a numerical coefficient estimated at the steady state. Hence, at the "leading order", all of the dependence on the machine is reduced to just a number, $\chi_0$, which can be determined. This is the real advantage of this approach. As an example of calculation, we have considered the simplest case of IGNITOR-plasmas. In this case, we solved the time differential equation for
$\tau_{E}$ where the parameters ({\i.e.}, the initial condition as well as the
coefficient appearing in the differential equation) have been estimated at the steady state. The solution of this equation is in agreement with the one obtained by solving numerically the dynamic balance
equations, with the aid of a transport model \cite{airoldi}.

\noindent In this work, we shall also justify the dynamic scaling laws, like
\begin{equation}\label{I2b}
\tau_{E}=C~I_{p}^{\alpha_{1}}\ {\bar{n}}_{e}^{\alpha_{2}}P^{a_{3}} M^{\alpha_4},
\end{equation}
\noindent where $C$ is a constant and $M$ is the effective mass, respectively (note that when the plasma is a mixture, due to the dependence of particle transport properties on particle mass and charge, $M$ is also time dependent). In particular, we shall prove that the dynamic expression for the energy confinement time, like Eq.~(\ref{I2b}), is solution of the differential equation for $\tau_{E}$, which can be obtained by combining Eq.~(\ref{I2a}) with the energy balance equation.

\noindent The paper is organized as follows. In Section~(\ref{tDE}), we show
that Eqs~(\ref{I2b}) satisfy a nonlinear differential equation of the second order in time, tacking into account the (experimentally
established) slow variation in time of the coefficient entering in this
equation. Successively, we show that this equation can also be derived from
the energy balance equation, combined with definition (\ref{I2a}). This will
allow a linking of the scaling coefficients with the (measurable) second time
derivatives of the heat power loss, which at the leading order may also be
estimated at the stationary state. These tasks will be accomplished in the
Section~(\ref{BDE}). As an example of an application, in the Section
(\ref{JETTO}), we compare the solution obtained by solving the differential
equation for the energy confinement time with the numerical simulations
obtained using the code JETTO \cite{airoldi}, for the specific case of
IGNITOR-plasmas. Concluding remarks can be found in Section (\ref{conclusions}).

\section{Differential Equation Satisfied by the ITER Scalings}

\label{tDE}

The expression for the energy confinement time, obtained by scaling laws,
raises several questions. Firstly, Eq.~(\ref{I1}) applies quite well to a
large number of Tokamaks (ASDEX, JET, DIII-D, ALCATOR C-Mod, COMPASS, etc.)
and it is currently used for predicting the energy confinement time for
Tokamaks, which are presently in construction (ITER) or will be constructed in
the future (DEMO). Hence, the first objective of this work is to understand
\textit{the main reason for such a "universal" validity}. Secondly, it is
legitimate to ask "\textit{where does this expression originate from} ?". More
concretely, "\textit{Is it possible to determine the (minimal) differential
equation which is satisfied by expression} (\ref{I2b}) ? ". In
case of a positive answer, "\textit{Is it possible to re-obtain this (minimal)
differential equation from the balance equations and, in particular, from the
energy balance equations} ?". Finally, "\textit{How can we estimate the values
of the scaling coefficients} $\alpha_{i}$ ?". In this Section, we shall
determine the (minimal) differential equation satisfied by Eqs~(\ref{I2b}). In the next Section we shall prove that, near the stationary
state, this differential equation can be re-obtained from the energy balance equation.

\noindent The equations of one-dimensional plasma dynamics, in toroidal
geometry, assuming the validity of the standard model, can be brought into the
form (see, for example, \cite{balescu})
\begin{align}
\label{tDE0a} &
\!\!\!\!\!\!\!\!\!\!\!\!\!\!\!\!\!\!\!\!\!\!\!\!\!\!\!\!\!\frac{\partial
n_{e}}{\partial t}=-\frac{1}{r}\frac{\partial}{\partial r}\Bigl(r<\gamma
_{r}^{e}>\Bigr)\nonumber\\
& \!\!\!\!\!\!\!\!\!\!\!\!\!\!\!\!\!\!\!\!\!\!\!\!\!\!\!\!\!\frac{3}{2}
\frac{\partial p}{\partial t}+\frac{1}{r}\frac{\partial}{\partial
r}\Bigl[r\Bigl(<q_{e}>+<q_{i}>+\frac{5}{2}(1+Z^{-1})T_{e}<\gamma_{r}
^{e}>\Bigr)\Bigr]=\nonumber\\
& \qquad\qquad\qquad\qquad\qquad\qquad\frac{c}{4\pi}\frac{E_{0}B_{0\phi}}
{Rr}\frac{\partial}{\partial r}\Bigl(\frac{r^{2}}{q(r)}\Bigr)+S_{gain-loss}
\end{align}
\noindent with $r$ and $q(r)$ denoting the radial coordinate and the safety
factor, respectively. $p$, $n_{e}$, $T_{e}$ and Z are the total plasma
pressure, the electron density, the electron temperature and the ion charge
number, respectively. Here, $<\cdots>$ denotes the surface-average operation.
$<q_{\zeta}>$ and $<\gamma_{r}^{e} >$ are the averaged radial heat flux of
species $\zeta$ ($\zeta=e$ for electrons and $\zeta=i$ for ions) and the
averaged electron flux, respectively. $c$ and $E_{0}$ are light speed and the
external electric field, respectively, and $S_{gain-loss}$ is the source term,
\textit{i.e.} the loss and energy gain. Eq.~(\ref{tDE0a}) must be completed
with the transport equations, \textit{i.e.} with the thermodynamic flux-force
relations, in order to close the plasma dynamical equations. The $0-D$ power
balance equation is now derived as follows. Eq.~(\ref{tDE0a}) is integrated
over the volume of the plasma and then divided by the plasma volume $V$. We
obtain
\begin{equation}
\label{tDE0b}\left\{
\begin{array}
[c]{ll}
& \!\!\!\!{\dot N}_{e}=-\Gamma\\
& \!\!\!\!{\dot W}_{e}+P_{Q}=P_{tot}
\end{array}
\right.
\end{equation}
\noindent with
\begin{align}
\label{tDE0c} & \!\!\!\!\!\!\!\!\!\!\!\!\!\!\!\!N_{e}\equiv V^{-1}\int n_{e}
dV\quad; \quad\Gamma\equiv V^{-1}\int\frac{1}{r}\frac{\partial}{\partial
r}\Bigl(r<\gamma_{r}^{e}>\Bigr)dV\\
& \!\!\!\!\!\!\!\!\!\!\!\!\!\!\!\!{W_{e}}\equiv\frac{3}{2}V^{-1}\int
pdV\ \ ;\ \ P_{tot}\equiv V^{-1}\int\Bigl[\frac{c}{4\pi}\frac{E_{0}B_{0\phi}
}{Rr}\frac{\partial}{\partial r}\Bigl(\frac{r^{2}}{q(r)}\Bigr)+S_{gain-loss}
\Bigr]dV\nonumber\\
& \!\!\!\!\!\!\!\!\!\!\!\!\!\!\!\!{P_{Q}}\equiv V^{-1}\int\Bigl(\frac{1}
{r}\frac{\partial}{\partial r}\Bigl[r\Bigl(<q_{e}>+<q_{i}>+\frac{5}
{2}(1+Z^{-1})T_{e}<\gamma_{r}^{e}>\Bigr)\Bigr]\Bigr)dV\nonumber
\end{align}
\noindent where the "dot" over the variables stands for the (total) time
derivative ($d/dt$).

\noindent The energy confinement time is defined as
\begin{equation}
\label{tDE1}\tau_{E}=\frac{W_{e}}{P_{tot}-{\dot W}_{e}}=\frac{W_{e}}{P_{Q}}
\end{equation}
\noindent From definition (\ref{tDE1}), we find
\begin{equation}
\label{tDE2}{\dot\tau}_{E}P_{Q}+\tau_{E}{\dot P}_{Q}-{\dot W}_{e}=0
\end{equation}
\noindent Note that the stationary state is reached when $P_{Q}=P_{tot}$.
Hence, at the steady state (corresponding to $t=t_{0}$) we have
\begin{equation}
\label{tDE3a}{\dot W}_{e}\mid_{t=t_{0}}\equiv{\dot W}_{e}^{0}=0
\end{equation}
\noindent At the steady state, we find
\begin{equation}
\label{tDE3}\tau_{E}(t_{0})\equiv\tau_{E}^{0}=\frac{W_{e}^{0}}{P^{0}_{tot}
}\quad;\quad\frac{d\tau_{E}}{dt}{\Bigg\arrowvert}_{t=t_{0}}
\!\!\!\!\!\!\!\!\!\!\equiv{\dot\tau}_{E}^{0}=0
\end{equation}
where $W_{e}^{0}$ and $P^{0}_{tot}$ indicate the values of $W_{e}$ and
$P_{tot}$, estimated at the steady state, respectively.

\noindent Eq.~(\ref{I2b}) may be re-written in the generic form:
\begin{equation}\label{tDE4}
\tau_{E}=C X_{1}^{\alpha_{1}}X_{2}^{\alpha_{2}}\cdots
X_{n}^{\alpha_{n}}
\end{equation}
\noindent where $X_{1}$, $X_{2}$, $\cdots$ are a positive and independent
system of variables $X_{i}$, and $\alpha_{i}$ the \textit{scaling parameters},
respectively. For simplicity, we firstly suppose that in Eq.~(\ref{tDE4}) all
the variables $X_{i}$ are time-dependent. The case whereby $X_{i}$ is a
collection of variables dependent on time, as well as variables not-dependent
on time, will be treated in the following sub-Section \textit{Analysis in the
Physics Variables}. Note that $C$ is a (dimensional) constant satisfying the
condition
\begin{equation}
\label{tDE5}C=\tau_{E}^{0} X_{1}(t_{0})^{-\alpha_{1}}X_{2}(t_{0})^{-\alpha
_{2}}\cdots X_{n}(t_{0})^{-\alpha_{n}}%
\end{equation}
\noindent Unless stated otherwise, in the sequel we shall adopt the summation
convention on the repeated indexes. By taking the logarithm of Eq.~(\ref{tDE4}
) we find
\begin{equation}
\label{tDE6}y=\log C +{\alpha_{1}}\xi_{1}+{\alpha_{2}}\xi_{2}+\cdots\alpha
_{n}\xi_{n}
\end{equation}
\noindent with $y\equiv\log\tau_{E}$ and $\xi_{i}\equiv\log X_{i}$ (with
$i=1,\cdots n$). The first and the second derivatives of $y$, with respect to
variable $\xi_{i}$, read respectively
\begin{equation}
\label{tDE7}\frac{\partial y}{\partial\xi_{i}}=\alpha_{i}\qquad;\qquad
\frac{\partial^{2} y}{\partial\xi_{i}\partial\xi_{j}}=0
\end{equation}
\noindent In terms of variable $\tau_{E}$, instead of $y$, we get
\begin{equation}
\label{tDE8}\frac{\partial\tau_{E}}{\partial\xi_{i}}=\tau_{E}\alpha_{i}
\qquad;\qquad\tau_{E}\frac{\partial^{2}\tau_{E}}{\partial\xi_{i}\partial
\xi_{j}}-\frac{\partial\tau_{E}}{\partial\xi_{i}}\frac{\partial\tau_{E}
}{\partial\xi_{j}}=0
\end{equation}
\noindent The differential equation with respect to time is easily obtained by
tacking into account the identities
\begin{equation}
\label{tDE9}{\dot\tau}_{E}=\frac{\partial\tau_{E}}{\partial\xi_{i}}{\dot\xi
}_{i}=\tau_{E}\alpha_{i}{\dot\xi}_{i}\quad;\quad\frac{\partial^{2}\tau_{E}
}{\partial\xi_{i}\xi_{j}}{\dot\xi}_{i}{\dot\xi_{j}}={\ddot\tau}_{E}
-\frac{\partial\tau_{E}}{\partial\xi_{i}}{\ddot\xi}_{i}={\ddot\tau}_{E}
-\tau_{E}\alpha_{i}{\ddot\xi}_{i}%
\end{equation}
\noindent By multiplying the second equation of Eqs~(\ref{tDE8}) by ${\dot\xi
}_{i}{\dot\xi}_{j}$ and by summing over indexes, we finally obtain \textit{the
differential equation satisfied by the ITER scaling laws}
\begin{equation}
\label{tDE10}\tau_{E}{\ddot\tau}_{E}-{\dot\tau}_{E}^{2}=\Bigl(\sum_{i=1}
^{n}\alpha_{i}{\ddot\xi}_{i}(t)\Bigr)\tau_{E}^{2}%
\end{equation}
\noindent Eq.~(\ref{tDE10}) should be solved with the initial conditions
(\ref{tDE3}):
\begin{equation}\label{tDE11}
\left\{
\begin{array}
[c]{ll}
& \!\!\!\!\tau_{E}{\ddot\tau}_{E}-{\dot\tau}_{E}^{2}=\chi(t)\tau_{E}^{2}\\
& \\
& \!\!\!\!\tau_{E}^{0}=\frac{W_{{e}_{}}^{0}}{P_{tot}^{0}}\quad;\quad{\dot\tau}
_{E}^{0}=0
\end{array}
\right.
\end{equation}
\noindent with $\chi(t)\equiv\Bigl(\sum_{i=1}^{n}\alpha_{i}{\ddot\xi}
_{i}(t)\Bigr)$. We have derived two differential equations for the time
derivatives of $\tau$, the first equation of Eq.~(\ref{tDE9}) which is first
order and also Eq.~(\ref{tDE10}) which is second order. It may appear hopeless
to solve these equations, as they depend on $\alpha_{i}{\dot\xi}_{i}(t)$ and
$\chi(t)=\alpha_{i}{\ddot\xi}_{i}(t)$ respectively, which in turn depend on
the full dynamics of the system. The critical fact which makes our approach
useful is that the second time derivatives of the logarithm of $X_{i}$ are
generally weakly dependent on time. As a result, one may approximate $\chi(t)$
to be a constant, $\chi_{0}$. In this sense, all of the dependence on the
machine is reduced to just a number, which can be determined. The evolution of
$\tau_{E}$ can then be obtained uniquely by integrating Eq.~(\ref{tDE10}) with
the initial conditions (\ref{tDE3}). Such an approach would not work for the
first equation of Eq.~(\ref{tDE9}) as $\alpha_{i}{\dot\xi}_{i}(t)$ depends
strongly on time, indeed it vanishes at the initial stationary state and then
becomes nonzero as the state evolves.

\noindent It is not difficult to check that the nonlinear equation
~(\ref{tDE11}) is the "minimal" differential equation, in the sense that
Eq.~(\ref{tDE11}) admits one, and only one, solution (\textit{i.e.}, the
nonlinear differential equation (\ref{tDE11}) \textit{does not generate} additional solutions).

\noindent It may appear hopeless to solve Eq.~(\ref{tDE11}), as it depends on the coefficient $\chi=\Bigl(\sum_{i=1}^{n}\alpha_{i}{\ddot\log X}_{i}\Bigr)$, which in turn depend on the full dynamics of the system. The critical fact which makes our approach useful is that the second time derivatives of the logarithm of $X_i$ are generally weakly time-dependent. In all the cases examined by the authors, $\chi(t)$ is very well approximated (numerically) by a linear function in time
\begin{equation}\label{tDE15}
\chi(t)=\simeq \chi_0(t-t_0)
\qquad\mathrm{with}\qquad \chi_0=-\frac{1}{t_0}\sum_{i=1}^{n}\alpha_{i}{\ddot\xi}_{i}(t_{0})
\end{equation}
\noindent Hence, all of the dependence on the machine is reduced to just a number, $\chi_0$, which can be \textit{estimated at the steady state}.

\vskip 0.2truecm

\section{Differential Equation for the Energy Confinement Time}

\label{BDE}

The aim of this Section is to obtain the differential equation for the energy
confinement time from the balance equations. In analogy with Eq.~(\ref{tDE11}%
), the coefficients of this differential equation should be expressed only in
terms of the internal energy $W_{e}$ and the total power $P_{tot}$. To this
end, let us reconsider the energy balance equation Eq.~(\ref{tDE0b}) and the
definition of the energy confinement time, Eq.~(\ref{tDE1}). Taking the
derivative of Eq.~(\ref{tDE2}) with respect to time, after a little algebra,
we get
\begin{equation}
\label{BDE1}\tau_{E}{\ddot\tau}_{E}-{\dot\tau}_{E}^{2}=-f(t)\tau_{E}%
^{2}-g(t)\tau_{E}{\dot\tau}_{E}
\end{equation}
\noindent with
\begin{align}\label{BDE3} 
& f(t)\equiv\frac{{\ddot P}_{tot}-{\dddot W}_{e}}{P_{tot}-{\dot
W}_{e}}-\frac{{\ddot W}_{e}}{W_{e}}=-\chi(t)\\
& g(t)\equiv\frac{{\dot P}_{tot}-{\ddot W}_{e}}{P_{tot}-{\dot W}_{e}}%
+\frac{{\dot W}_{e}}{W_{e}}\nonumber
\end{align}
\noindent Note that the dimensions of $f(t)$ and $g(t)$ are $[t]^{-2}$ and
$[t]^{-1}$, respectively. Finally, the \textit{differential equation for the
energy confinement time} reads
\begin{equation}
\label{BDE4}\left\{
\begin{array}
[c]{ll}
& \!\!\!\!\tau_{E}{\ddot\tau}_{E}-{\dot\tau}_{E}^{2}+f(t)\tau_{E}^{2}%
+g(t)\tau_{E}{\dot\tau}_{E}=0\\
& \\
& \!\!\!\!\tau_{E}^{0}=\frac{W_{{e}_{}}^{0}}{P_{tot}^{0}}\quad;\quad{\dot\tau}%
_{E}^{0}=0
\end{array}
\right.
\end{equation}
\noindent We might object that the previous equation has the same degree of
difficulty as the initial expression, Eq.~(\ref{tDE1}). However, as we shall
see more in detail in the next Subsection, the coefficients $g(t)$ and $f(t)$
possess special properties: close to the steady state $g(t)$ tends to vanish
and $f(t)$ is a function varying very slowing in time. So, at the leading
order, $g(t)\approx0$ and $f(t)$ may be estimated at the stationary state [see
Eq.~(\ref{tDE15}) and the discussion after Eq.~(\ref{tDE11})]. This is the
real advantage of Eq.~(\ref{BDE4}) with respect to Eq.~(\ref{tDE1}) :
\textit{Eq.~(\ref{BDE4}) allows determining the dynamic behaviour of the
energy confinement time when the system is close to the steady state, solely
by the knowledge of one coefficient estimated at the stationary state}.
Moreover, from the previous Section we know that this equation admits one (and
only one) solution corresponding to the ITER scalings. A concrete application
of Eq.~(\ref{BDE4}) can be found in the Section~(\ref{JETTO}). Note that
Eq.~(\ref{BDE4}) may be re-written in the more convenient form
\begin{equation}
\label{BDE4a}\left\{
\begin{array}
[c]{ll}
& \!\!\!\!{\dot\tau}_{E}=\tau_{E}y\\
& \!\!\!\!{\dot y}+g(t)y+f(t)=0\\
& \!\!\!\!\tau_{E}(t_{0})=\frac{W_{{e}_{}}^{0}}{P_{tot}^{0}}\quad;\quad y(t_{0})=0
\end{array}
\right.
\end{equation}
\noindent showing that the differential equation for the energy confinement
time may be expressed as two \textit{quasi-decoupled} differential equations
of first order in time derivative. The general solution of Eqs~(\ref{BDE4a})
may be brought into the form
\begin{equation}
\label{BDE4b}\tau_{E}(t)=\tau_{E}^{0}\exp\Bigl[-{\int_{t_{0}}^{t}%
}\!\!dx^{\prime\prime}\Bigl(\exp\bigl(-\!\!\!\int_{t_{0}}^{x^{\prime\prime}%
}\!\!\!\!dx\ \!g(x)\bigr) \bigl[\int_{t_{0}}^{x^{\prime\prime}}%
\!\!\!dx^{\prime}f(x^{\prime})\exp\bigl(\int_{t_{0}}^{x^{\prime}%
}\!\!\!dxg(x)\bigr)\bigr] \Bigr)\Bigr]
\end{equation}
\noindent By taking into account that $f(t)=-\sum_{i=1}^{n}\alpha_{i}{\ddot
\xi}_{i}$ (with $\xi_{i}=\log X_{i}$), solution~(\ref{BDE4b}) generalizes the
ITER scaling laws out of the steady state, reducing to Eq.~(\ref{tDE4}) close
to the stationary state. Eq.~(\ref{BDE4b}) shows that close to the steady state, the leading contribution to the mathematical expression for the energy confinement time is provided by the power laws. However, when we deviate from the steady state, supplementary contributions, which are different from the power ones, may modify the mathematical form of the power laws significantly. Generally, for ITER, these contributions tend to lower the numerical value of the energy confinement time.

\vskip0.2truecm \noindent$\bullet$ \textbf{Differential Equation for the
Energy Confinement Time Near the Steady State} \vskip0.2truecm \noindent The
term $P_{tot}$ is specified as follows
\begin{equation}
\label{BDE5}P_{tot}=P_{\alpha}(T)-P_{b}(T)+P_{Aux}(r,t)
\end{equation}
\noindent where $P_{\alpha}(T)$ is the alpha power, $P_{b}(T)$ is the power
radiation loss

\noindent(Bremsstrahlung) and $P_{aux}$ is the external heating power density
supplied to the system (\textit{e.g.} ohmic heating power or external RF),
respectively. The alpha power and the Bremsstrahlung power loss depend
explicitly on the temperature of the plasma. The auxiliary heating power is
operational during both the transient and steady states. This is the dominant
source of external heating power, and it is assumed to be deposited in the
plasma with a known profile, independent of $p$ and $T$. Hence, $P_{Aux}%
=P_{Aux}(r,t)$. The time derivative of $P_{tot}$ reads
\begin{equation}
\label{BDE6}{\dot P}_{tot}=\frac{\partial P_{\alpha}}{\partial T}{\dot
T}-\frac{\partial P_{b}}{\partial T}{\dot T}+{\dot P}_{Aux}%
\end{equation}
\noindent At the steady state ${\dot T}(t_{0})=0$ and ${\dot P}_{Q}%
(t_{0})={\dot P}_{Aux}(t_{0})=0$. Consequently, from the energy balance
equation we find that also ${\ddot W}_{e}(t_{0})=0$. By taking into account
Eqs~(\ref{tDE3a}) and (\ref{BDE3}), we get $g(t)\rightarrow0$ as the system
approaches the steady state. Hence, near the stationary state, we find
\begin{equation}
\label{BDE7}\left\{
\begin{array}
[c]{ll}
& \!\!\!\!\tau_{E}{\ddot\tau}_{E}-{\dot\tau}_{E}^{2}\simeq\chi(t)\tau_{E}%
^{2}\\
& \\
& \!\!\!\!\tau_{E}^{0}=\frac{W_{{e}_{}}^{0}}{P_{tot}^{0}}\quad;\quad{\dot\tau}%
_{E}^{0}=0
\end{array}
\right.
\end{equation}
with
\begin{equation}
\label{BDE8}\chi(t)=-\frac{{\ddot P}_{tot}-{\dddot W}_{e}}{P_{tot}}=\sum
_{i=1}^{n}\alpha_{i}{\ddot\xi}_{i}(t)\simeq\chi_0(t-t_0)
\end{equation}
\noindent where Eq.~(\ref{tDE15}) has been used. As shown in the Section
(\ref{tDE}), Eq.~(\ref{BDE7}) admits one (and only one) solution,
corresponding to the ITER scalings Eq.~(\ref{I2b}). Note that Eq.~(\ref{BDE8})
provides the desired relation between the exponent coefficients $\alpha_{i}$
and the macroscopic quantities $P_{tot}$ and $W_{e}$. If we have $n$ free
exponent coefficients $\alpha_{i}$, we can set the following $n$ relations
\begin{equation}
\label{BDE9}\sum_{i=1}^{n}\alpha_{i}{\ddot\xi}_{i}(t_{k})=-\frac{{\ddot
P}_{tot}(t_{k})-{\dddot W}_{e}(t_{k})}{P_{tot}(t_{k})}\quad\mathrm{with}\quad
k=0,1,\cdots, n-1
\end{equation}
\noindent Eq.~(\ref{BDE9}) link the exponent coefficients with variables
which, at least in principle, are under the control of the experimental
physicist. \vskip0.2truecm \noindent$\bullet$ \textbf{Analysis in the
"Physics" Variables}
\vskip0.2truecm

\noindent As mentioned, Eqs~(\ref{I1}) and (\ref{I2}) are composed by several
variables independent of time (\textit{e.g.}, major and minor radii,
elongation etc.). In this case, it is more convenient to express the energy
confinement time only in terms of the time-dependent variables. Let us suppose
that $m$ variables are time-dependent and the remaining $n-m$ not. In this
case, the energy confinement time takes the form [see Eq.~(\ref{tDE5})]
\begin{equation}
\label{BDE10}\tau_{E}=\tau_{E}^{0}\Bigl(\frac{X_{1}^{\alpha_{1}}}%
{X_{1}^{\alpha_{1}}(t_{0})}\Bigr)\Bigl(\frac{X_{2}^{\alpha_{2}}}{X_{2}%
^{\alpha_{2}}(t_{0})}\Bigr)\cdots\Bigl(\frac{X_{n}^{\alpha_{m}}}{X_{m}%
^{\alpha_{m}}(t_{0})}\Bigr)
\end{equation}
\noindent where, now, the independent variables $X_{i}^{\alpha_{i}}%
(t)/X_{i}^{\alpha_{i}}(t_{0})$ are dimensionless. Note that in this case
variables $\xi_{i}$ are defined as $\xi_{i}=\log(X_{i}/X_{i}(t_{0}))$ (no
summation convention over the repeated indexes). Of course, this operation
reduces the number of independent variables. However, this number may be
reduced further if, instead of "engineering variables", the confinement time
is expressed in terms of "physics" parameters such as $\rho^{\star}$
(normalized Larmor radius), $\beta$ (normalized pressure), $\nu^{\star}$
(collisionality), etc. Indeed, according to the observation of Kadomtsev, the
transport in the plasma core should be fundamentally governed by three
physical dimensionless plasma parameters $\rho^{\star}$, $\beta$ and
$\nu^{\star}$ \cite{kadomtsev}. In this respect, an interesting paper is
Ref.~\cite{martin}. In \cite{martin} the authors show that, due to the
Kadomtsev constraint, the final expression for the ELMy H-mode thermal
confinement time has only one free exponent coefficient, according to the
law:
\begin{equation}
\label{BDE11}\tau_{E}^{best}=2\pi\ 10^{-3} I_{p}\epsilon^{-1}n_{e}%
^{\alpha_{n_{e}}}{P^{\star}}^{-(6+8\alpha_{n})/15}%
\end{equation}
\noindent with $P^{\star}$ denoting the density of the power loss
(\textit{i.e.}, $P^{\star}\equiv P/V$). With the choice $\alpha_{n_{e}}=1/2$,
in "physics" variables, scaling (\ref{BDE11}) goes as $\alpha_{\rho^{\star}%
}=-1$ (\textit{i.e.} a gyro-Bohm-like scaling), $\alpha_{\beta}=-0.5$ and
$\alpha_{\nu^{\star}}=0$. This choice may be tested by using Eq.~(\ref{BDE9})
which, in this particular case, reads
\begin{equation}
\label{BDE12}15\alpha_{n_{e}}{\ddot{\log\!n}}_{0e}-(6+8\alpha_{n_{e}}%
){\ddot{\log\!P}}_{0}^{\star}=-15\frac{{\ddot P}^{0}_{tot}-{\dddot W}_{0e}%
}{P^{0}_{tot}}%
\end{equation}
\noindent where Eq.~(\ref{BDE10}) has been taken into account. We find
\begin{equation}
\label{BDE13}\alpha_{n_{e}}=\frac{6P^{0}_{tot}{\ddot{\log\!P}}_{0}^{\star
}-15\bigl({\ddot P}_{0tot}-{\dddot W}_{0e}\bigr)}{P^{0}_{tot}\bigl(15{\ddot
{\log\!n}}_{0e}-8{\ddot{\log\!P}}_{0}^{\star}\big)}%
\end{equation}

\bigskip

\section{Comparison with the Numerical Simulation of the Balance Equations for
an L-mode Tokamak-plasma}

\label{JETTO}

As an example application, we consider in this Section the case of one of the
simplest L-mode Tokamak-plasma where the evolution of the energy confinement
time has been estimated by solving numerically the balance equations,
completed with a transport model. In \cite{airoldi} we find the profile of
$\tau_{E}$ against time for Ignitor-plasma. The numerical solution has been
obtained by using the code JETTO. To compare this profile with the numerical
solution of Eq.~(\ref{tDE11}), we should firstly estimate $t_{0}$, $\tau
_{E}^{0}$ and $\chi_0=\frac{1}{t_0}{\ddot P}_{Q}(t_{0})/P_{tot}(t_{0})$ [see Eq~(\ref{tDE15}) and
(\ref{BDE3})]. In \cite{cardinali}, we have estimated the values of these
parameters for Ignitor subject to ICRH power (\textit{i.e.}, $P_{Aux}
=P_{ICRH}$). The scenario is considered where IGNITOR is led to operate in a
slightly sub-critical regime by adding a small fraction of ${}^{3}He$ to the
nominal $50-50$ Deuterium-Tritium mixture. The difference between power lost
and alpha heating is compensated by an additional ICRH power equal to 1.46 MW,
which should be able to increase the global plasma temperature via collisions
between ${}^{3}He$ minority and the background $D-T$ ions. The analytical
expression for the ICRH power profiles inside the plasma has been deduced by
fitting the numerical results giving an expression for $P_{Aux}=P_{ICRH}(r)$,
which is essentially independent of the bulk temperature. Denoting the ICRH
power-density as $P^{\star}_{ICRH}$, we have
\begin{equation}
\label{JETTO1}P^{\star}_{ICRH}(r)=P^{\star}_{0ICRH}\exp\bigl[{\tilde\alpha
}_{2} B(r_{ICRH})/B_{0\phi}\bigr]\exp\bigl[-(r-r_{ICRH})^{2}/\Delta\bigl]
\end{equation}
\noindent with $P^{\star}_{0ICRH} = 6.59126\ 10^{-6} MW / m^{3}$,
${\tilde\alpha}_{2}= 15.3478$ and $\Delta= 0.0477032$, respectively.

\begin{figure}[ptb]
\hfill\begin{minipage}[t]{.45\textwidth}
\begin{center}
\hspace{-0.5cm}
\resizebox{1.0\textwidth}{!}{%
\includegraphics{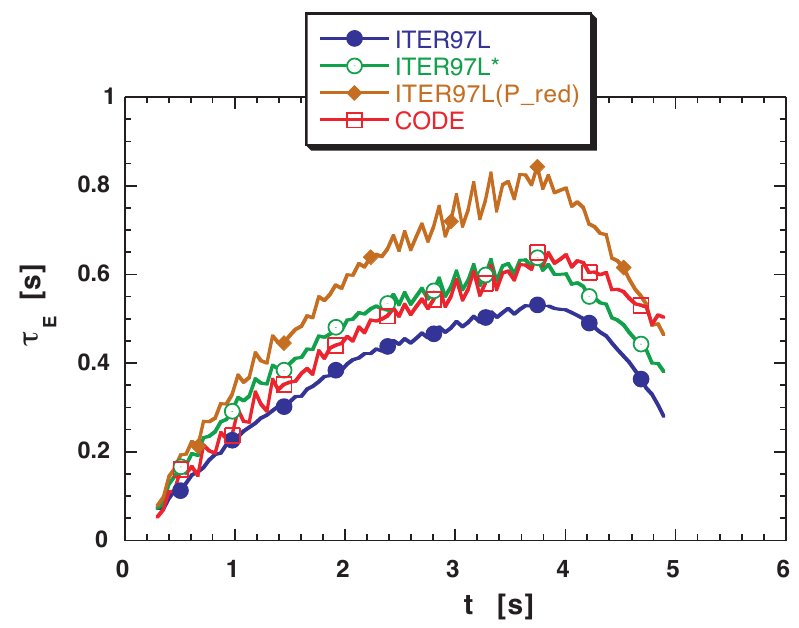}
}
\caption{
{\bf This is a reproduction of the picture, which appears in} \cite{airoldi}. {\it Energy confinement time evolution estimated in \cite{airoldi} by solving with JETTO the balance equations (completed with a transport model): ITER97L scaling (full dots - blue line), $ITER97L^{\star}$ scaling (open dots - green line) and ITER97L(P\_red) scaling (brown line)}.
}
\label{a1}
\end{center}
\end{minipage}
\hfill\begin{minipage}[t]{.45\textwidth}
\begin{center}
\hspace{-0cm}
\resizebox{0.98\textwidth}{!}{%
\includegraphics{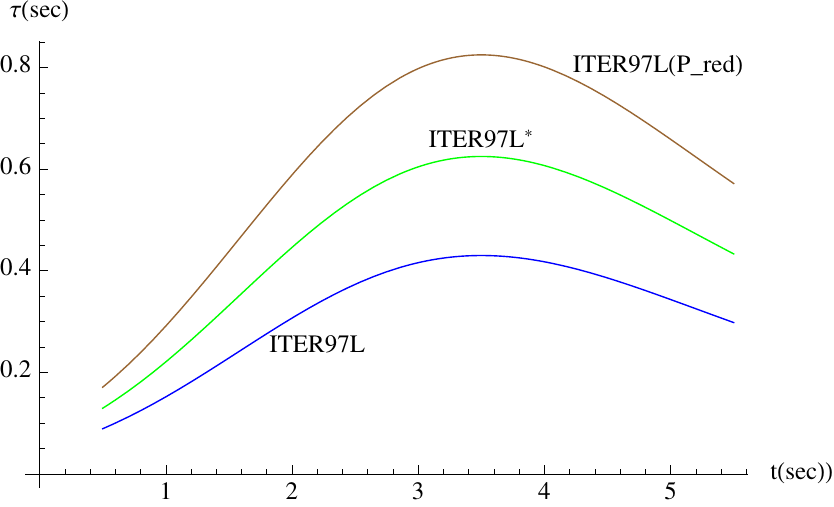}
}
\caption{
{\it Solutions of Eq.~(\ref{tDE11}) at the three values of ($\tau_E^0, t_0$). Blue line: ($t_0, \tau_E^0$)=(0.35sec, 0.43sec) (ITER97L ), Green line:  ($t_0, \tau_E^0$)=(0.35sec, 0.625sec) ($ITER97L^{\star}$) and Brown line: ($t_0, \tau_E^0$)=(0.35sec, 0.825sec)= (ITER97L(P\_red))}.
}
\label{s1}
\end{center}
\end{minipage}
\hfill\end{figure}

\noindent The value of $\tau_{E}^{0}$ has been estimated by the expression
\cite{cardinali}
\begin{equation}
\label{JETTO2}\tau_{E}^{0}=\frac{12n_{e}T}{E_{\alpha}n_{e}^{2}<\sigma
v>_{D-T}-C_{B}n_{e}^{2}T^{1/2}+4P^{\star}_{ICRH}}%
\end{equation}
\noindent with $E_{\alpha}$ and $C_{B}$ denoting the energy at which the alpha
particles are created ($3.5 MeV$), and the Bremsstrahlung constant,
respectively. $\sigma$ is the reaction cross section giving a measure of the
probability of a fusion reaction as a function of the relative velocity of the
two reactant nuclei. $<\sigma v>_{D-T}$ provides an average over the
distributions of the product of cross section and velocity $v$. In the core of
the plasma we found \cite{cardinali} $\tau_{E}^{0}=0.43sec$, $t_{0}%
=3.5sec$ and $\chi_0=0.171429sec^{-3}$.
Figs~(\ref{a1}) reports on the energy confinement time, $\tau_{E}$, against
time for Ignitor-plasmas in the above mentioned conditions. The profiles have
been obtained by solving (with the code JETTO) the balance equations and refer
to the ITER scalings $ITER97L$ (full dots), $ITER97L^{\star}$ (open dots) and
ITER97L \cite{airoldi}. Fig.~(\ref{s1}) shows the solutions of the
differential equation for the ITER scalings, Eq.~(\ref{tDE11}), at the three
values of ($t_{0}, \tau_{E}^{0}$): ($t_{0}, \tau_{E}^{0}$)=(0.35sec, 0.43sec)
(ITER97L - blue line), ($t_{0}, \tau_{E}^{0}$)=(0.35sec, 0.625sec)
($ITER97L^{\star}$ - green line) and ($t_{0}, \tau_{E}^{0}$)=(0.35sec,
0.825sec)= (ITER97L(P\_red) - brown line).
\noindent Note that in \cite{airoldi} the authors evaluate the ITER scalings
by using the reduced power $P_{red}=P_{tot}-P_{RadTot}$, whereas in our work
we use $P_{tot}$, which includes the Bremsstrahlung radiation loss. This may
explain the little difference between the numerical \cite{airoldi} and the
analytical slopes.

\section{Conclusions}

\label{conclusions}

A large database on plasma energy confinement in Tokamaks can be summarized in
single empirical value of $\tau_{E}$, referred to as the
\textit{ITER-scalings}. These expressions are "Universal", in the sense that
they apply to a large number of Tokamaks. Scalings are expressed in terms of
product of powers of independent variables [see Eq.~(\ref{tDE4})] and
correspond to the $L$-mode as well as the $H$-mode confinements. The
recommended scaling for ITER operation remains the $IPB98$ scaling law, while
this issue is further investigated. In this work we have shown that the ITER
scalings satisfy a general non-linear differential equation of second order in
time. The value provided by the database for ITER scaling laws, coincides with
$\tau_{E}^{0}$, estimated by Eq.~(\ref{I2a1}), with $W_{estat.}$ and
$P_{Qstat.}$ evaluated by solving the stationary balance equations. To
estimate the dynamic confinement time, we determined the differential equation
for $\tau_{E}$ by combining the energy balance equation with definition
(\ref{I2a}). We found Eqs~(\ref{BDE4}). We have solved this equation by taking
into account that, in vicinity of the steady state, the coefficient $g(t)$
tends to vanish and, at the leading order, $f(t)$ is (almost) a constant independent of
time, which may be evaluated at the stationary
state. This is the real advantage of the proposed approach: {\it close to the steady state, the differential equation for the energy confinement time} $\tau_E$ {\it reduces to} 
\begin{equation}
\left\{
\begin{array}
[c]{ll}
& \!\!\!\!\tau_{E}{\ddot\tau}_{E}-{\dot\tau}_{E}^{2}=\chi(t)\tau_{E}^{2}\\
& \\
& \!\!\!\!\tau_{E}^{0}=\frac{W_{{e}_{}}^{0}}{P_{tot}^{0}}\quad;\quad{\dot\tau}
_{E}=0\nonumber
\end{array}
\right.
\end{equation}
\noindent {\it where "at the leading order"} $\chi(t)$ {\it is a numerical constant, which may be estimated at the stationary state}. As a result, one may approximate $\chi(t)$ to be a constant, $\chi_0$ or better, by a linear function $\chi(t)=\chi_0(t-t_0)$. In this sense, all of the dependence on the machine is reduced to just a number, $\chi_0$, which can be estimated at the steady state. Far from the stationary state the differential equation for $\tau_{E}$ contains a nonlinear extra term, which behaves as $\sim\!\tau_{E}{\dot\tau}_{E}$. This extra term tends to modify the mathematical form of the power laws. For ITER, the main effect of this nonlinear extra term is to lower the numerical value of the energy confinement time. The general solution is given by Eq.~(\ref{BDE4b}), which reduces to the one admitting the
ITER scaling power laws as the system approaches the steady state. We have also seen
that the scaling coefficients may be linked to the variables which, at least
in principle, are under the control of the experimental physicist. The
validity of our approach has been tested by analyzing a concrete example of
Tokamak-plasma where the profile of the energy confinement time has been
previously determined by solving the balance equations (with the
\textit{auxilium} of a transport model). The solution of the differential
equation for the ITER scaling is in a fairly agreement with the numerical finding.

\section{Acknolwedgments}

\label{Ack}

JE is supported by NSFC MianShang grant 11375201.

\bigskip

\begin{thebibliography}{99}                                                                                               
\bibitem {kardaun1}O.J.W.F. Kardaun, \textit{Classical methods of statistics:
with applications in fusion-oriented plasma physics}, \textit{Springer
Science} \& \textit{Business} ISBN 3540211152 (2005).

\bibitem {tsunematsu}T. Tsunematsu, \textit{Fusion Engineering and Design},
\textbf{15}, Issue 4, 309 (1991).

\bibitem {doyle} E.J. Doyle, W.A. Houlberg, Y. Kamada, V. Mukhovatov, T.H. Osborne, A. Polevoi, G. Bateman, J.W. Connor, J.G. Cordey, T. Fujita, X. Garbet, T.S. Hahm, L.D. Horton, A.E. Hubbard, F. Imbeaux, F. Jenko, J.E. Kinsey, Y. Kishimoto, J. Li, T.C. Luce, Y. Martin, M. Ossipenko, V. Parail, A. Peeters, T.L. Rhodes, J.E. Rice, C.M. Roach, V. Rozhansky, F. Ryter, G. Saibene, R. Sartori, A.C.C Sips, J.A. Snipes, M. Sugihara, E.J. Synakowski, H. Takenaga, T. Takizuka, K. Thomsen, M.R. Wade, H.R. Wilson, ITPA Pedestal, \textit{Nucl. Fusion},
\textbf{47}, S18 (2007).

\bibitem {wagner}F. Wagner, \textit{The Physics Basis of ITER Confinement},
2nd ITER Int. Summer School (Kyushu University, Japan, 2009) (New York: AIP)
\textit{AIP Conf. Proc.}, \textbf{31}, 1095 (2009).

\bibitem {ITER}ITER \textit{Physics Expert Groups on Confinement and Transport
and Confinement Modelling and Database}, ITER Physics Basis Editors and ITER
EDA, Naka Joint Work Site, Mukouyama, Naka-machi, Naka-gun, Ibaraki-Ken, Japan\textit{Nucl. Fusion}, IAEA, Vienna, \textbf{39}, No 12, 2137 (1999).

\bibitem {luce}T.C. Luce, C.C. Petty, and J.G. Cordey, \textit{Plasma Phys.
Control. Fusion}, \textbf{50}, 4, 043001 (2008).

\bibitem {Kadomtsev}B.B. Kadomtsev, \textit{Sov. J. Plasma Phys.}, \textbf{1},
295 (1975).

\bibitem {kritz}A.H.Kritz, J.Kinsey, T.Onjun, I.Voitsekhovich, G.Bateman,
R.Waltz, G.Staebler, \textit{Burning Plasma Projections with Internal
Transport Barriers}, ITPA Meeting on Burning Plasma Transport, NIFS, Tokio
(Japan), 10-12 September 2001.

\bibitem {weiland}J.Weiland, \textit{Predictive Simulations of ITER-FEAT
Performance}, 28th EPS Conference, Madeira, P2.039 (2001).

\bibitem {bateman}G. Bateman, A. H. Kritz, T.Onjun and A. Pankin,
\textit{Private communication}, 7 Dec., 2001.

\bibitem {cordey}J.-G. Cordey, \textit{Plasma Phys. Control. Fusion},
\textbf{39}, B115 (1997).

\bibitem {fusionwiki}\textit{Scaling Law}, FusionWiki, jointly hosted by
\textit{LNF} and {FuseNet}. It is associated with two domains:
fusionwiki.ciemat.es (http://fusionwiki.ciemat.es/wiki/Scaling\_law) and wiki.fusenet.eu.

\bibitem {kardaun}O. Kardaun, \textit{Nucl. Fusion}, \textbf{42}, 841 (2002).

\bibitem {cordey1}
J.G. Cordey, J.A. Snipes, M. Greenwald, L. Sugiyama, O. J. W. F. Kardaun, F. Ryter, A. Kus, J. Stober, J. C. DeBoo, C. C. Petty, G. Bracco, M. Romanelli, Z. Cui, Y. Liu, J.G. Cordey, K. Thomsen, D. C. McDonald, Y. Miura, K. Shinohara, K.Tsuzuki, Y. Kamada, T. Takizuka, H. Urano, M. Valovic, R. Akers, C. Brickley, A. Sykes, M.J. Walsh, S.M. Kaye, C. Bush, D. Hogewei, Y. Martin, A. Cote, G. Pacher, J. Ongena, F. Imbeaux, G.T. Hoang, S. Lebedev, A. Chudnovskiy, V. Leonov, \textit{IAEA 20th Fusion Energy Conference}, paper IAEA-CN-116/IT/P3-32, Vilamoura, Portugal, (2004).

\bibitem {yushmanov} P.N. Yushmanov, T. Takizuka, K.S. Riedel, O.J.W.F. Kardaun, J.G. Cordey, S.M. Kaye and D.E. Post, \textit{Nucl. Fusion}, \textbf{30}, 1999 (1990).

\bibitem {kaye} S.M. Kaye, M. Greenwald, U. Stroth, O. Kardaun, A. Kus, D. Schissel, J. DeBoo, G. Bracco, K. Thomsen, J.G. Cordey, Y. Miura, T. Matsuda, H. Tamai, T. Takizuda, T. Hirayama, H. Kikuchi, O. Naito, A. Chudnovskij, J. Ongena and G. Hoang, \textit{Nucl. Fusion}, \textbf{37}, 1303 (1997).

\bibitem {dinklage} A. Dinklage, H. Maa$\beta$berg, R. Preuss, Yu.A. Turkin, H. Yamada, E. Ascasibar, C.D. Beidler, H. Funaba, J.H. Harris, A. Kus, S. Murakami, S. Okamura, F. Sano, U. Stroth, Y. Suzuki, J. Talmadge, V. Tribaldos, K.Y. Watanabe, A. Werner, A. Weller and M. Yokoyama, \textit{Nucl. Fusion},
\textbf{47}, 9, 1265 (2007).

\bibitem {yamada} H.Yamada , J.H.Harris , A.Dinklage , E.Ascasibar , F.Sano , S.Okamura , U.Stroth , A.Kus , J.Talmadge , S.Murakami , M.Yokoyama , C.D.Beidler , V.Tribaldos , K.Y.Watanabe, \textit{31st EPS Conference on Plasma Phys. London} 28 June - 2 July 2004, ECA Vol. \textbf{28G}, 5.099 (2004).

\bibitem {cardinali}A. Cardinali and G. Sonnino, \textit{Analysis of the
Thermonuclear Instability including Low-Power ICRH Minority Heating in
IGNITOR}, submitted for publication in EPJD (2014).

\bibitem {airoldi}A. Airoldi and G. Cenacchi, \textit{Nucl. Fusion},
\textbf{41}, No. 6, 687 (2001).

\bibitem {balescu}R. Balescu, \textit{Transport Process in plasmas - Vol. II},
Elsevier Science Publication, North-Holland, (1988).

\bibitem {kadomtsev}B.B. Kadomtsev, \textit{Sov. J. Pl. Phys.}, \textbf{1},
295, (1975).

\bibitem {martin}O. Sauter and Y. Martin, \textit{Nucl. Fusion}, \textbf{40},
No. 5, 955 (2000).
\end{thebibliography}
\end{document}